\documentclass{svproc}
\usepackage{url}

\usepackage{amsmath}
\usepackage{algorithm}
\usepackage{algpseudocode}
\usepackage{float}
\usepackage{color}
\usepackage{graphicx}

\newcommand{\Schrodinger}{Schr\"{o}dinger }\newcommand{\cm}{$\rm{cm}^{-1}$}
\newcommand{\amorse}{a_{\rm{Morse}}}

\begin{document}
\mainmatter              

\title{Low temperature scattering with the R-matrix method: the Morse potential}
\titlerunning{R-matrix scattering: the Morse potential} 
\author{Tom Rivlin$^1$ \and Laura K. McKemmish$^{1,2}$
 \and Jonathan Tennyson$^1$}

\authorrunning{Tom Rivlin et al.} 
%
\tocauthor{Tom Rivlin, Laura K. McKemmish, Jonatnan Tennyson}
\institute{$^1$ Department of Physics and Astronomy, University College London,\\ London, WC1E~6BT, UK,\\
$^2$ School of Chemistry, University of New South Wales, Kensington, Sydney, Australia\\
\email{t.rivlin@ucl.ac.uk and j.tennyson@ucl.ac.uk}}

\maketitle              

\begin{abstract}

Experiments are starting to probe collisions and chemical reactions
between atoms and molecules at ultra-low temperatures. We have developed
a new theoretical procedure for studying these collisions using the R-matrix
method. Here this method is tested for the atom -- atom collisions
described by a Morse potential. Analytic solutions for continuum
states of the Morse potential are derived and compared with numerical
results computed using an R-matrix method where the inner region wavefunctions
are obtained using a standard nuclear motion algorithm. Results are given for
eigenphases and scattering lengths. Excellent agreement
is obtained in all cases. Progress in developing a general procedure
for treating ultra-low energy reactive and non-reactive collisions
is discussed.
\keywords{Low temperature, elastic scattering, R-matrix, Morse potential}

\end{abstract}

\section{Introduction}
\label{Introduction}

The ability to perform very low-energy collisions between heavy particles
is leading to a quiet revolution at the border between atomic physics
and experimental quantum chemistry \cite{14StHuYe}. Studies of reactive
and non-reactive collisions at temperatures very significantly below 1 K
are starting to probe processes which are not easily resolved at higher
temperatures.   These experiments study chemical reactions
and scattering at the quantum scattering limit where, asymptotically,
only a few partial
waves contribute 
\cite{12QuJuxx}. 

To address these problems theoretically requires the development of
new computational techniques. Recently, we proposed adapting R-matrix
theory to the study of ultra-low energy reactive and non-reactive, heavy-particle
collisions \cite{jt643}. R-matrix theory involves the division
of space into an inner region encompassing the whole collision complex
and an outer region where species involved in the scattering can be
separately identified. Procedures based on the computable R-matrix method
have proved outstandingly successful for the study of electron collisions
with atoms and molecules \cite{11Burke.Rmat,jt474}, and are increasingly
being adopted in other areas \cite{10DeBa.Rmat}. In the computable 
R-matrix method, the Schr\"odinger equation for the restricted inner
region is solved once and for all for each scattering symmetry, independent
of the precise scattering energy. For heavy particle scattering this
procedure is particularly appropriate for reactive or
non-reactive collisions which occur over deep potential
energy wells. Thus, for example, H + H$_2$ collisions do not occur over
a deep well as the H$_3$ system is only weakly bound \cite{jt11},
while collisions between H$^+$ + H$_2$  occur over the deep well of the
H$_3^+$ potential energy surface \cite{jt535}.

Strongly bound systems with deep potential energy wells support many
bound states. Even the very lowest continuum states which are
associated with ultra-low energy scattering feel the effect of these
many bound states which lie below them in energy.  The result is that even
the lowest scattering state has a 
complicated wavefunction which couples many channels which are
asymptotically closed. It is well-known that this situation leads to a
plethora of quasibound states, or resonances, in the near-dissociation
region \cite{89CaMcxx.H3+,13MaQuGo,jt494,13SzCsxx,jt443}. Use of the
R-matrix method allows the region of the deep potential well to be treated
using variational nuclear motion programs which are capable of giving
highly accurate results for energy-independent problems with
complicated wavefunctions \cite{jt512}.  It is then only necessary to
treat a few partial waves in the energy-dependent outer region. In this region
it may be necessary to propagate solutions to very large interparticle
separations \cite{15LaJaAo} and to scan over the many energies necessary
to characterise narrow resonances.

At present we are in the process of developing a heavy particle
R-matrix scattering code, RmatReact, based on the use of a variety of
variational nuclear motion codes in the inner region
\cite{jt609,jt338,jt339,Trove,Trove2}. Doing this involves developing
computational procedures which extend methods of the solutions into the
continuum \cite{13SzCsxx,jt253,jt443}. In particular,
 the problem must be solved within a
finite region and, critically, use basis functions which give reliable
amplitudes at the R-matrix boundary.  These amplitudes, and the
associated inner region energies, are used to construct the
scattering energy-dependent R-matrix which links the inner and outer regions \cite{jt643}.

In this paper we report on tests we have performed using our
methodology for the Morse oscillator potential. Section 2 gives an
overview of the general theory while Section 3 demonstrates
that the scattering problem can be solved analytically for a
Morse oscillator potential. This allows the rigorous assessment of our
numerical procedures, which are discussed in Section 4. Results are given in Section 5, and conclusions and some pointers to our future work are given in
the final Section.

\section{Theory: The RmatReact method}
\label{TheRMatrixMethod}

The theory behind the RmatReact method has been discussed extensively
\cite{jt643,11Burke.Rmat,jt474}, and much of this explanation
derives from those discussions. The general principle behind the
method is the partitioning of space into an inner and outer region,
dependent on the reaction coordinate, as discussed above.

In the case of two atoms colliding there is only one reaction
coordinate: the internuclear distance $r$. A point $r=a_0$ is defined
such that any internuclear distance lower than that is the inner
region and any distance larger is the outer region. 

Within the inner region, the system is treated as a bound diatom, and
the eigenenergies and eigenfunctions of the radial \Schrodinger
equation with the Morse potential can be determined using software
built for nuclear motion calculations. Because the eigenfunctions and
values refer to the bound states, they are independent of scattering
energy. Likewise, in the outer region, the system is treated as a pair of weakly interacting, unbound atoms. Each atom will have associated atomic channels describing its quantum state.

The inner region was solved in this work using a discrete variable
representation (DVR) \cite{00LiCaxx.methods} grid method based on the Lobatto shape
functions, which have the property of always having a point defined on both boundaries of the grid. Manolopoulos and Wyatt
\cite{93Manolopoulos.Rmat,88MaWy.Rmat} pioneered the use of these functions
for scattering problems.  Lobatto shape functions \cite{16WeissteinLobQuad.Rmat} can be used to obtain
simple expressions for the components of the Hamiltonian matrix,
making it computationally efficient to diagonalise whilst avoiding
much of the expensive integration usually involved in constructing a
Hamiltonian matrix.
Once the inner region has been solved to obtain a diagonalised Hamiltonian matrix, a matrix known as the R-matrix, can be constructed on the boundary $a_0$. 
The R-matrix is constructed from the scattering energy, $E$, the bound eigenenergies, and the values of the eigenfunctions on the boundary $a_0$, known as the \textit{surface amplitudes}.

For a given angular momentum quantum number $J$, if the $m^{\rm{th}}$ surface amplitude associated with the $i^{\rm{th}}$ atomic channel is defined as $w_{im}^J(a_0)$, the $m^{\rm{th}}$ eigenenergy is defined as $E_m^J$, and the scattering wavefunction for atomic channel $i$ is defined to be $F_i^J(r,E)$, then the R-matrix has two equivalent definitions at $a_0$:

\begin{equation}
\label{eq:Rmatrix1}
F_i^J(a_0,E) = \sum_{j=1}^{N_{\rm{ch}}^J} a_0 R_{ij}^J(a_0,E) \frac{dF_j^J(r,E)}{dr}\bigg|_{r=a_0},
\end{equation}
\begin{equation}
\label{eq:Rmatrix2}
R_{ij}^J(a_0,E) = \frac{\hbar^2}{2\mu a_0}\sum_{m=1}^{N}\frac{w_{im}^J(a_0)w_{jm}^J(a_0)}{E_m^J - E},
\end{equation}
where the sum in Eq.~(\ref{eq:Rmatrix1}) is over the $N_{\rm{ch}}^J$ atomic channels for a given value of $J$ considered in the scattering event, and the sum in Eq.~(\ref{eq:Rmatrix2}) is over the $N$ solutions to the \Schrodinger equation within an atomic channel.

In the $J=0$, single channel case considered in this work, $i = j$ and
Eq.~(\ref{eq:Rmatrix1}) reduces down to a single term, $N_{\rm{ch}} =
1$, and the single R-matrix element is defined as $R(a_0,E)$.
Furthermore, $w_{im}^J(a_0)$ becomes a single surface amplitude, and
the sum is over the $N$ surface amplitudes.

As Eq.~(\ref{eq:Rmatrix1}) suggests, the R-matrix can be thought of as
the `log-derivative' of the channel function $F^J(r,E)$, which is an
outer region function. However Eq.~(\ref{eq:Rmatrix2}) shows that
the R-matrix can be constructed as a sum over the eigenfunctions and
energies of the inner region.  The fact that these inner and outer
region definitions of the R-matrix are equivalent is what gives the
R-matrix method its value: information about the energy-independent inner
region provide the starting point for
obtaining scattering information.

In the outer region, it is assumed that the potential is small and
slowly varying, compared to the deep wells of the inner region. As
such, it is possible to use methods which iteratively solve the
\Schrodinger equations over finite distances to \textit{propagate} the
R-matrix from the boundary at $a_0$ to an asymptotic distance $a_p$.
At $a_p$, the potential is assumed to be zero. In this work, the
propagation method due to Walker and Light \cite{wl80,11Burke.Rmat} is
used.

For the single-channel, the propagation algorithm takes as
its input the R-matrix element at the inner region boundary,
$R(a_0,E)$, and produces the R-matrix element at the asymptotic
distance, $R(a_p,E)$. To produce the value of the R-matrix at the
outer region point $a_s$, $R_s(a_s, E)$, the iteration equation takes as its input the value of the R-matrix at $a_{s-1}$, $R_{s-1}(a_{s-1},E)$, 
and has the form:
\begin{equation}
\label{eq:LightWalker2}
R_s = \frac{-1}{a_s \lambda_s}\left( \frac{1}{\tan(\lambda_s \Delta a)} + \frac{2}{\sin(2 \lambda_s \Delta a)} \left( a_{s-1} R_{s-1} \lambda_s \tan(\lambda_s \Delta a) - 1\right)^{-1} \right),
\end{equation}
where $\Delta a = a_s - a_{s-1}$, and 
\begin{equation}
\label{eq:LambdaDefinition2}
\lambda_s^2 = \frac{2\mu}{\hbar^2}\left(E - V(a_s) \right).
\end{equation}

At the asymptotic distance $a_p$, the R-matrix is used to construct
the K-matrix using the equation \cite{11Burke.Rmat}:

\begin{equation}
K_{ij}^J(k) = -\left(\frac{s^J(k r) - R_{ij}^J(a_p,E) k r s^{J'}(k r)}{c^J(k r) + R_{ij}^J(a_p,E) k r c^{J'}(k r)}\right),
\label{eq:KmatrixNumerical}
\end{equation}
where
\begin{subequations}
\label{eq:sphericalBessels}
\begin{align}
& s^{\nu}(x) = x j^{\nu}(x)\label{eq:sphericalBesselss}\\
& c^{\nu}(x) = -x y^{\nu}(x),\label{eq:sphericalBesselsc}
\end{align}
\end{subequations}
where $j^{\nu}(x)$ is the Spherical Bessel Function of the First Kind, $y^{\nu}(x)$ is the Spherical Bessel Function of the Second Kind, and $s^{J'}(x)$ and $c^{J'}(x)$ are the derivatives with respect to x of $s^{J}(x)$ and $c^{J}(x)$ respectively.

In the single-channel, Eq.~(\ref{eq:KmatrixNumerical}) reduces  to

\begin{equation}
K(k) = \frac{R(a_p,E) k r - \tan{k r}}{1 + R(a_p,E) k r \tan{k r}}.
\label{eq:KmatrixNumerical2}
\end{equation}

\begin{figure}[h]
	\centering
	\includegraphics[scale=0.31]{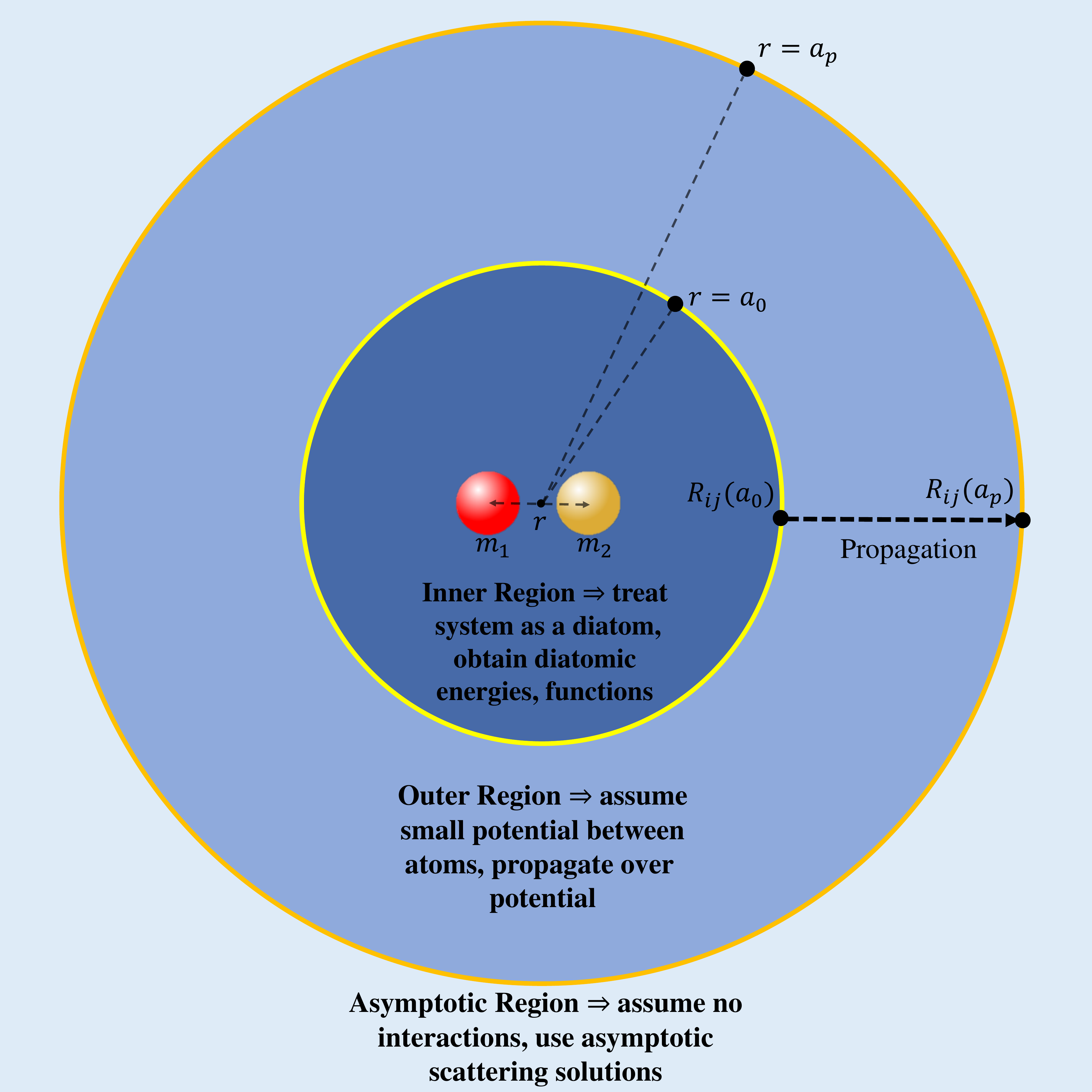}
	\caption{Schematic outlining the partitioning of space into an inner, outer and asymptotic region in the R-matrix method.}
\label{fig:rmatschemat}
\end{figure}

The inner, outer and asymptotic regions are illustrated in Fig.~(\ref{fig:rmatschemat}).

\section{Analytic Scattering in Morse Oscillators}
\label{AnalyticScattering}

\subsection{Morse Oscillator Solutions} 
When there is no angular momentum and hence no centrifugal term, the Morse potential for a diatom as a function of the internuclear distance $r$ has the algebraic form:

\begin{equation}
\label{eq:Morse}
V(r) = D_e \left(\left(1 - e^{- \amorse (r - r_e)}\right)^2-1\right),
\end{equation}
where $D_e$ is the well depth (assuming the zero of potential energy is placed at the dissociation energy), $r_e$ is the equilibrium position, or position of the well minimum, and $\amorse$ is a scaling parameter (the so-called Morse parameter) affecting the shape of the well.

The analytic eigenfunctions and eigenenergies of the \Schrodinger equation with a Morse potential are well known. For the radial time-independent
\Schrodinger equation
\begin{equation}
\label{eq:Schrodinger2}
\left( -\frac{\hbar^2}{2m}\frac{d^2}{dr^2} + D_e e^{-2\amorse(r - r_e)} - 2D_e e^{-\amorse(r-r_e)}\right) \Psi_n = E_n \Psi_n,
\end{equation}
the bound eigenenergies $E_n$ and eigenfunctions $\Psi_n$ are given by \cite{29Morse.Rmat}:
\begin{equation}
\label{eq:MorseEnergies}
E_n^{\rm Morse} = -D_e + 2\amorse \sqrt{\frac{D_e \hbar^2}{2\mu}}\left( n + \frac{1}{2} \right) - \frac{1}{4D_e}\left( 2\amorse\sqrt{\frac{D_e \hbar^2}{2\mu}}\left( n + \frac{1}{2}\right) \right)^2,
\end{equation}
and
\begin{equation}
\label{eq:MorseFunctions}
\Psi_n^{\rm Morse} = N_n z^{\left(1/(\amorse r_0)-n-1/2\right)}\exp{\left(\frac{-z}{2}\right)}L_n^{\left(2/(\amorse r_0)-2n-1\right)}(z),
\end{equation}
where $L_n^{(\alpha)}(z)$ is the $n^{\rm th}$ associated Laguerre polynomial, and $N_n$ is a normalising factor given by
\begin{equation}
\label{eq:Nn}
N_n = \left(\frac{\left(\frac{2}{\amorse r_0}-2n - 1 \right) \amorse \Gamma\left( n+1 \right)}{\Gamma\left( \frac{2}{\amorse r_0}-n \right)}\right)^{\frac{1}{2}},
\end{equation}
where $\Gamma(x)$ is the standard Gamma function.

\subsection{Scattering Observables}
It is possible to derive analytic scattering observables for a quantum
scattering event involving the Morse oscillator potential energy curve
because the time-independent \Schrodinger equation with a Morse
potential is analytically soluble. 

Similar to Eq.~(\ref{eq:Schrodinger2}), for a scattering event between particles with reduced mass $\mu$ with energy $E$ interacting over a Morse potential, the radial wavefunction $\Psi(r)$ is given by the time-independent radial \Schrodinger equation: 

\begin{equation}
\label{eq:Schrodinger1}
\left( -\frac{\hbar^2}{2\mu}\frac{d^2}{dr^2} + D_e e^{-2\amorse(r - r_e)} - 2D_e e^{-\amorse(r-r_e)}\right) \psi = E \psi.
\end{equation}

By defining 
\begin{equation}
\label{eq:kequation}
k=\sqrt{\frac{2\mu E}{\hbar^2}},
\end{equation}
one constraint that may be placed on $\psi(r)$ is that in the
no-potential limit it must behave like the wavefunction of a free
particle, i.e. it must be a plane wave. Likewise, this means that in
the infinite distance limit where the potential's strength tends to
zero, the wavefunction must be sinusoidal such that:
\begin{equation}
\label{eq:FunctionAsymptote}
\lim_{r\to\infty} \psi(r) = \sin(kr+\delta(k)),
\end{equation}
where $\delta(k)$ is defined to be the phase shift (also known as the eigenphase) induced in the particle by its interaction with the potential.

Furthermore, by defining
\begin{equation}
\label{eq:rnought}
r_0 = \sqrt{\frac{\hbar^2}{2\mu D_e}},
\end{equation}
\begin{equation}
\label{eq:zequation}
z(r) = \frac{2}{\amorse r_0}e^{-a(r-r_e)},
\end{equation}
and
\begin{equation}
\label{eq:Phiequation}
\Phi(z) = z^{\frac{1}{2}}\psi(z),
\end{equation}
then it can be shown \cite{02RaMeNg.Rmat} that  Eq.~(\ref{eq:Schrodinger1}) can be re-written as
\begin{equation}
\label{eq:Whittaker1}
\frac{d^2\Phi}{dz^2}+ \left( -\frac{1}{4} + \frac{1}{\amorse r_0 z} + \frac{\frac{1}{4} + \left(\frac{k}{\amorse}\right)^2 }{z^2} \right)\Phi(z) = 0.
\end{equation}
In this form, the equation is equivalent to the well-known Whittaker equation, whose solutions are the Whittaker functions. There are two linearly independent solutions to Eq.~(\ref{eq:Whittaker1}):
\begin{equation}
\label{eq:WhittakerFunctions}
\psi_{\pm}(z) = e^{-z/2}z^{\pm ik/\amorse}{}_1F_1\left(\frac{1}{2}-\frac{1}{\amorse r_0}\pm\frac{ik}{\amorse},1\pm\frac{2ik}{\amorse};z\right),
\end{equation}
where ${}_1F_1(x,y;z)$ is the Kummer confluent hypergeometric function of the first kind, and the $\psi_{\pm}(z)$ functions represent incoming and outgoing waves.

Using the results for the analytic scattering wavefunctions of the Morse potential in Eq.~(\ref{eq:WhittakerFunctions}), it is possible to construct an analytic equation for the eigenphase $\delta(k)$ associated with scattering  with the Morse potential. The eigenphase of the scattering event is desired because it can be used to generate other observables such as the cross section and scattering length.

The derivation below follows that of Rawitscher \textit{et al.} \cite{02RaMeNg.Rmat} and Selg \cite{16Selg.Rmat,12Selg.Rmat}.

The general solution $\psi(r)$ to Eq.~(\ref{eq:Schrodinger1}) can be written in terms of the two solutions to Eq.~(\ref{eq:Whittaker1}), which are given by Eq.~(\ref{eq:WhittakerFunctions}), such that:
\begin{equation}
\label{eq:AsymptoticWavefunction}
\psi(r) = C_{+} \psi_{+}(r) + C_{-} \psi_{-}(r),
\end{equation}
where $C_{\pm}$ are two constants. 

There are two boundary conditions on $\psi(r)$ that can be used to obtain an expression for the eigenphase. Firstly, the asymptotic radial function must vanish at $r=0$, such that $\psi(0) = 0$. This fact can be used to express one of the $C_{\pm}$ coefficients in terms of the other. Secondly the $r\to\infty$ asymptotic limit is given by Eq.~(\ref{eq:FunctionAsymptote}). As $r\to\infty$, $z\to0$. This means that due to a property of the Kummer confluent hypergeometric functions, both hypergeometric functions tend to $1$ as $r\to\infty$.

The S-matrix can be defined in the $r\to\infty$ limit as the negative of the ratio of the coefficients of the outgoing plane wave component of the asymptotic radial wavefunction to the incoming plane wave component \cite{11Burke.Rmat}. 

Then, by defining $z_0$ such that 
\begin{equation}
\label{eq:znought}
z(r=0) = z_0 = \frac{2}{\amorse r_0}e^{ar_e},
\end{equation}
the following expression can be obtained:
\begin{equation}
\label{eq:planewave}
\left( \frac{z}{z_0} \right)^{\pm\frac{ik}{\amorse}} = e^{\mp ikr}.
\end{equation}

Using the boundary conditions and Eq.~(\ref{eq:planewave}), one can obtain an expression for the ratio of the coefficients of $\psi_{\pm}$ in this limit, and hence one can obtain an analytic expression for the S-matrix:

\begin{equation}
\label{eq:S-matrix}
S(k) = \lim_{r\to\infty} \frac{C_+}{C_-} = \frac{{}_1F_1\left( \frac{1}{2} - \frac{1}{\amorse r_0} + \frac{ik}{\amorse}, 1 + \frac{2ik}{\amorse};z_0 \right)}{{}_1F_1\left( \frac{1}{2} - \frac{1}{\amorse r_0} - \frac{ik}{\amorse}, 1 - \frac{2ik}{\amorse};z_0 \right)}.
\end{equation}
Besides Eq.~(\ref{eq:FunctionAsymptote}), another way of defining the eigenphase is as the argument of the S-matrix, such that:
\begin{equation}
\label{eq:SmatrixEigenphase}
S(k) = e^{2i\delta(k)}.
\end{equation}
Note that the factor of $2$ in  the exponent is arbitrary, and other authors define it differently, depending on whether the eigenphase is defined as the argument of the S-matrix (as in \cite{16Selg.Rmat}), or as the arctangent of the K-matrix, 
which is equivalent to defining the eigenphase to be half of the argument of the S-matrix (as in this work, and Ref. \cite{02RaMeNg.Rmat}).

The analytic expression for the eigenphase is then given by:

\begin{equation}
\label{eq:AnalyticMorse}
\delta(k) = \frac{1}{2}\arg\left(\frac{{}_1F_1\left( \frac{1}{2} - \frac{1}{\amorse r_0} + \frac{ik}{\amorse}, 1 + \frac{2ik}{\amorse};z_0 \right)}{{}_1F_1\left( \frac{1}{2} - \frac{1}{\amorse r_0} - \frac{ik}{\amorse}, 1 - \frac{2ik}{\amorse};z_0 \right)}  \right).
\end{equation}

Once the eigenphase has been obtained for a given Morse potential, then many scattering observables can be derived, including the K-matrix, and the T-matrix (also known as the transition matrix):
\begin{equation}
\label{eq:K-matrix_1}
K(k)=\tan{\delta(k)},
\end{equation}
\begin{equation}
\label{eq:S-matrix_2}
S(k) = \frac{1-iK(k)}{1+iK(k)},
\end{equation}
\begin{equation}
\label{eq:T-Matrix_1}
T(k) = S(k) - 1.
\end{equation}
Note that other authors use different definitions of the T-matrix such as  the negative of its definition given here.

The total cross section at a given energy, $\sigma_{\rm tot}(k)$, which is the integral of the differential cross section over all solid angles,
 can be obtained from the eigenphase:
\begin{equation}
\label{eq:Total_Cross_Section}
\sigma_{\rm tot}(k) = \frac{4\pi}{k^2}\sin^2(\delta(k)).
\end{equation}

Finally the scattering length, $A$, and the effective range, $r_{\rm eff}$, are characteristic length scales associated with low-energy scattering. $A$ is defined as the limit
\begin{equation}
\label{scattering}
A = \lim_{k\to 0}\left(\frac{-\tan(\delta(k))}{k}\right),
\end{equation}
for the $J=0$, s-wave (lowest energy) eigenphase \cite{11Burke.Rmat}. The scattering length can be thought of as the low-energy $k \rightarrow 0$ limit of the gradient of the eigenphase. The effective range can be analytically determined through an integral over all space of the difference between the zero-energy scattering wavefunction, and the zero-energy potential-free scattering wavefunction \cite{49Bethe.Rmat}. It can be thought of as a length parameter which measures the overall effect the potential has on the scattering event, since it is defined by the difference between scattering in the cases with and without a potential. As such, calling it the effective \textit{range} of the potential is natural.

One way of obtaining these two quantities from the eigenphase is by taking a Taylor expansion of the eigenphase close to zero scattering energy \cite{11Burke.Rmat}:

\begin{equation}
\label{eq:ScatteringLengthAndEffectiveRange}
k\cot{\delta(k)} = \frac{-1}{A} + \frac{1}{2}r_{\rm{eff}}k^2 + O(k^4).
\end{equation}

\section{Method}
\subsection{Potentials Investigated} 
\begin{figure}   
\centering
\includegraphics[width=0.7\linewidth]{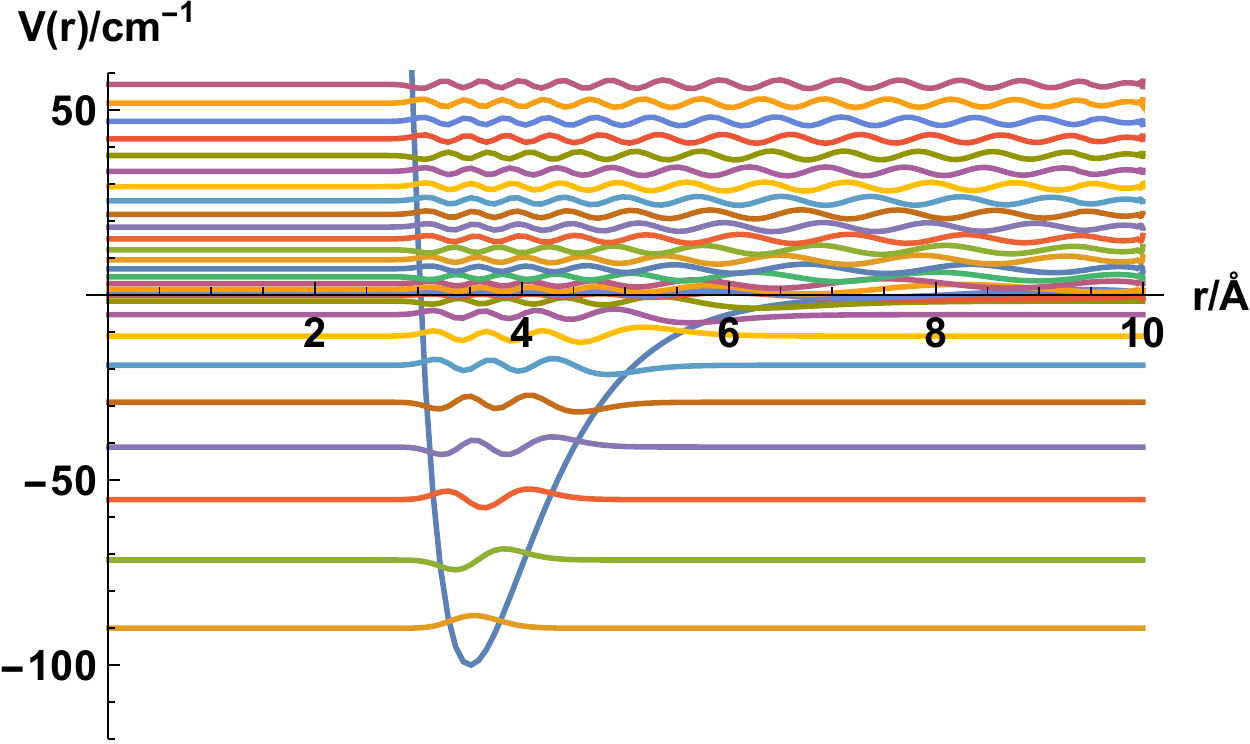}
\caption{A Morse oscillator potential energy curve for an $\rm{Ar}_2$-like potential with $D_e=100$~\cm , $r_e=3.5$~\AA, $a_{\rm{Morse}}=1.451455517$~\AA$^{-1}$. Wavefunctions of the vibrational bound states are also shown at their associated eigenenergies, along with with the continuum states between $0$ and $60$~\cm\ . The bound and continuum states were generated by solving the \Schrodinger equation with $\mu = 33.71525621$~Da with an R-matrix method with a boundary of $10$~\AA.}
\label{fig:morsepotential2} 
\end{figure}

The main Morse potential used in this work is presented in Fig.~(\ref{fig:morsepotential2}). This Morse potential uses parameters with reduced mass of $\mu = 33.71525621$~Da (and a value of $\hbar$
obtained from Qiang and Dong \cite{12MoTaNe.Rmat}). The value for $\mu$ was chosen for numerical convenience when testing the algorithm, as it meant that $\hbar^2/2\mu$ had a value of $0.5$ to seven decimal places in the units of \cm\ and \AA\ used in this work. 
The specific value used for $\amorse$ was chosen such that the ground state eigenenergy was 90~\cm\ to six decimal places, for ease of comparison. The values of $D_e$ and $r_e$ used in this work were chosen in analogy with the $\rm{Ar}_2$ dimer, which is currently being used to investigate the application of this method to more sophisticated potentials. The analytic eigenenergies were generated from these parameters and Eq.~(\ref{eq:MorseEnergies}).

\begin{figure}   
\centering
\includegraphics[width=0.7\linewidth]{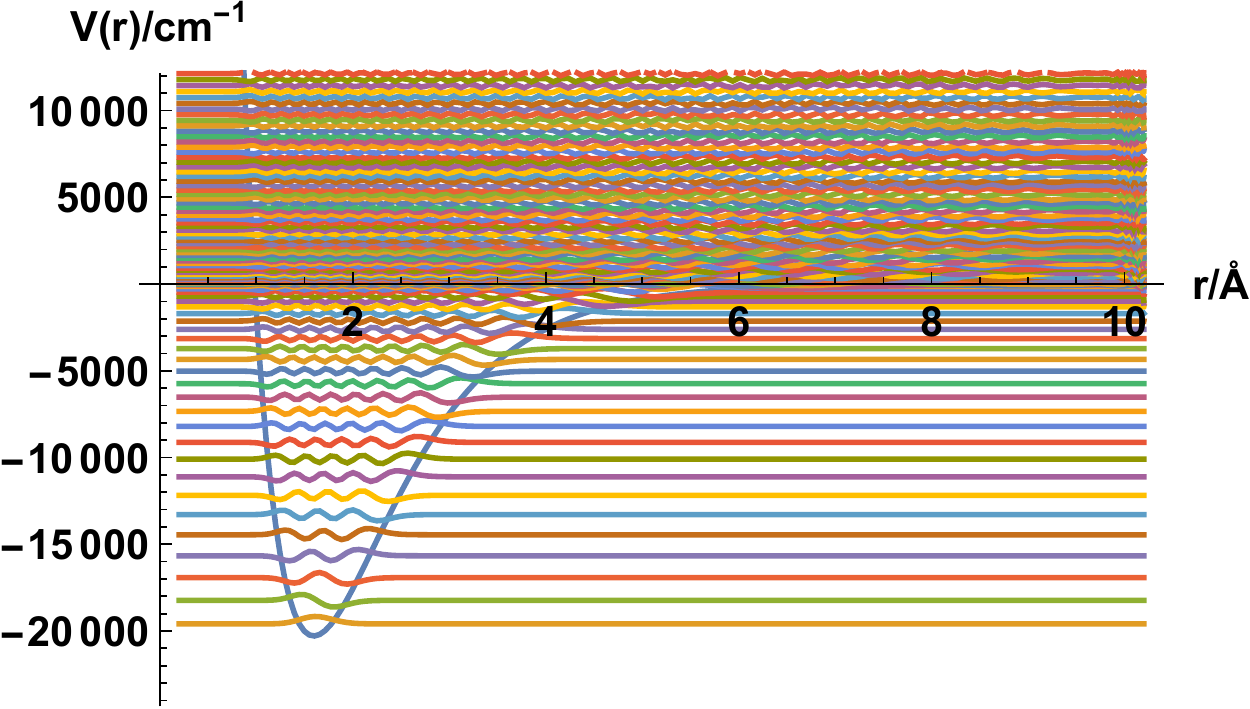}
\caption{Morse oscillator potential and states for LiH. Parameters used are $D_e = 20287.62581$~\cm , $r_e = 1.5956$~\AA , $a_{\rm{Morse}} = 1.128$~\AA$^{-1}$ \cite{07QiDo.Rmat}. The states were generated by solving the \Schrodinger equation with $\mu = 0.8801221$~Da \cite{07QiDo.Rmat} with an R-matrix method with a boundary of $10$~\AA.}
\label{fig:LiHPotential} 
\end{figure}

Other Morse potentials were tested, notably several obtained from \cite{07QiDo.Rmat} for actual diatoms: LiH, $\rm{H}_2$, HCl, and CO. 
 Fig.~(\ref{fig:LiHPotential}) shows one of these potentials: LiH. In this paper, we present only results for the Morse potential shown in Fig.~(\ref{fig:morsepotential2}). Similar numerical behaviour was observed for all of the potentials tested, however.

\subsection{Numerical Details}

The R-matrix method was used to generate scattering results, including the eigenphase and the scattering length, for the single-channel, $J=0$ Morse oscillator potential. These results are compared with the analytic results quoted above. 

In the construction of the R-matrix, the inner region bound system was solved numerically to generate the bound eigenenergies and radial eigenfunctions of two particles interacting over a Morse potential well. 
To generate the numeric results, $N = 200$ grid points and eigenfunctions were used to obtain the inner region eigenenergies (and amplitudes) using the Lobatto shape functions DVR method outlined in section \ref{TheRMatrixMethod}. The inner region was defined to range from $r_{\rm{min}} = 0.01$\AA\ to $a_0 = 10.0$\AA. 

The R-matrix was then constructed on the boundary and propagated to an
asymptotic radius. For the results presented in the following, the propagation was performed from $a_0 = 10.0$~\AA\ to
$a_p = 25.0$~\AA, with $N_{\rm{prop}} = 2500$ iterations of the propagation equation
over a uniform grid. The propagated R-matrix was then used to
construct the eigenphase for the $J=0$ Morse scattering event.

To explore the low-energy behaviour of the numeric method, the analytic
and numeric eigenphases were used to generate the 
scattering length and effective range. This was done by fitting the low-energy
plot to the form given in Eq.~(\ref{eq:ScatteringLengthAndEffectiveRange}) using
\textit{Mathematica}'s FindFit function over the lower scattering energy range
$k = 0.0004$~\AA\ to $k = 0.001$~\AA. (This is equivalent to $E = 8.0\times
10^{-8}$~\cm\ to $E = 5.0\times 10^{-7}$~\cm\ for this system.)


\section{Results}
\label{RMatrixResults}

\subsection{Comparison between analytic and numerical RmatReact results}
\label{ComparisonToAnalyticResults}

\begin{figure}   
\centering
\includegraphics[width=0.7\linewidth]{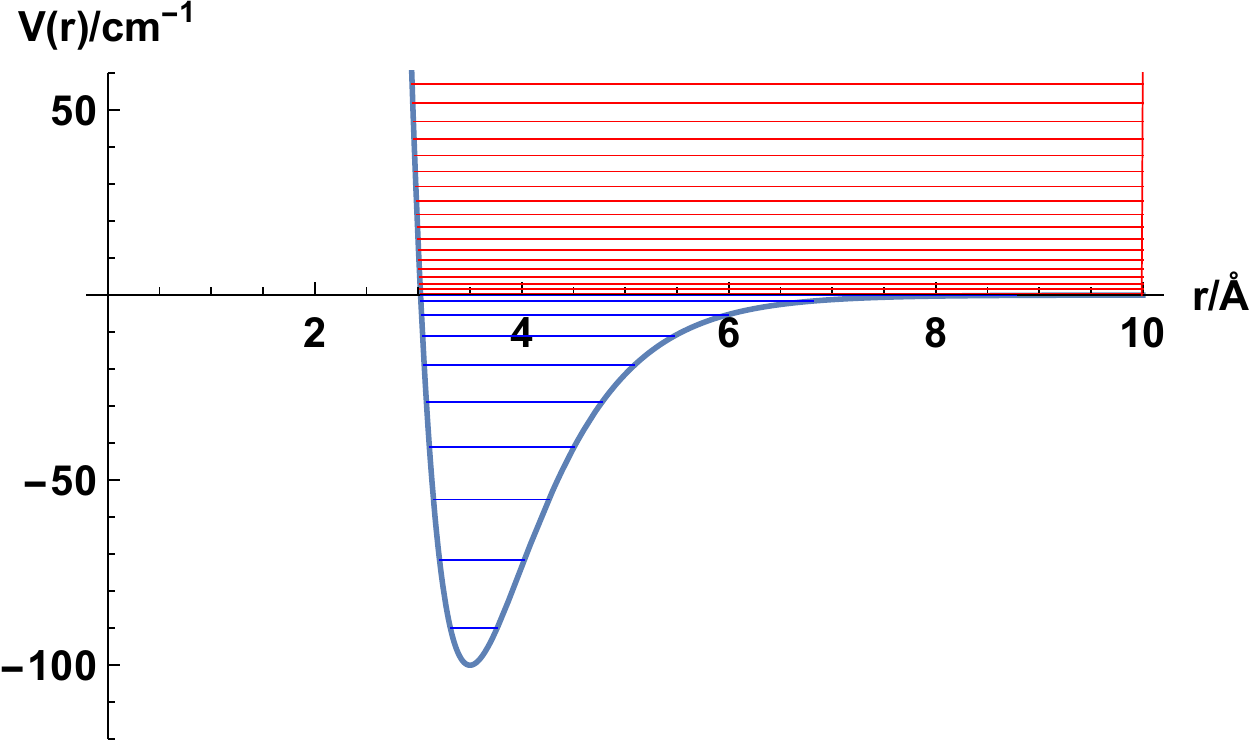}
\caption{The same Morse oscillator potential as in Fig.~(\ref{fig:morsepotential2}). Energy levels of the continuum states generated by the R-matrix below below $60$ \cm\ are coloured differently to the vibrational bound states in order to distinguish the states close to dissociation from the states just above dissociation. The R-matrix inner region boundary, $a_0 = 10$~\AA, is also highlighted.}
\label{fig:morsepotential3} 
\end{figure}

The numerical and analytic results for the eigenenergies are presented
in Table \ref{tab:EigenenergiesComparison}.  For low-lying states
whose wavefunctions are essentially completely contained in the inner
region, the agreement between the two methods is excellent. The final
two states are more diffuse, as seen in
Fig.~(\ref{fig:morsepotential2}), and hence they are more likely to
have significant amplitude outside the inner region. Due to this, the inner region
solution energies lies slightly below the true answer.

\begin{table}[]
\renewcommand{\arraystretch}{1.15}
\centering
\caption{Comparison of the analytic and numeric bound eigenenergies of the Morse diatomic system for vibrational energy levels $n=0$ to $9$. The relative error refers to the difference between each level's numeric and analytic values, divided by the analytic value (analytic minus numeric, divided by analytic).}
\label{tab:EigenenergiesComparison}
\begin{tabular}{llll}
\hline\hline
$n$  & Analytic  / \cm & R-matrix  / \cm & Relative error              \\ \hline
$0$  & $-90.000000$                 & $-90.000000$                 & $1.73\times 10^{-12}$  \\ 
$1$  & $-71.580042$                 & $-71.580042$                 & $4.59\times 10^{-11}$  \\ 
$2$  & $-55.266807$                 & $-55.266807$                 & $1.30\times 10^{-11}$    \\ 
$3$  & $-41.060295$                 & $-41.060295$                 & $1.82\times 10^{-11}$   \\ 
$4$  & $-28.960506$                 & $-28.960506$                 & $5.73\times 10^{-12}$  \\ 
$5$ & $-18.967441$                 & $-18.967441$                 & $1.33\times 10^{-12}$  \\ 
$6$  & $-11.081099$                 & $-11.081099$                 & $3.25\times 10^{-12}$  \\ 
$7$  & $-5.3014807$                 & $-5.3014807$                 & $-6.42\times 10^{-12}$ \\ 
$8$  & $-1.6285853$                 & $-1.6286033$                 & $-0.000011$           \\ 
$9$ & $-0.062413189$               & $-0.094633937$               & $-0.516$               \\ \hline\hline
\end{tabular}
\end{table}

\begin{figure}[H]
\includegraphics[scale=0.8]{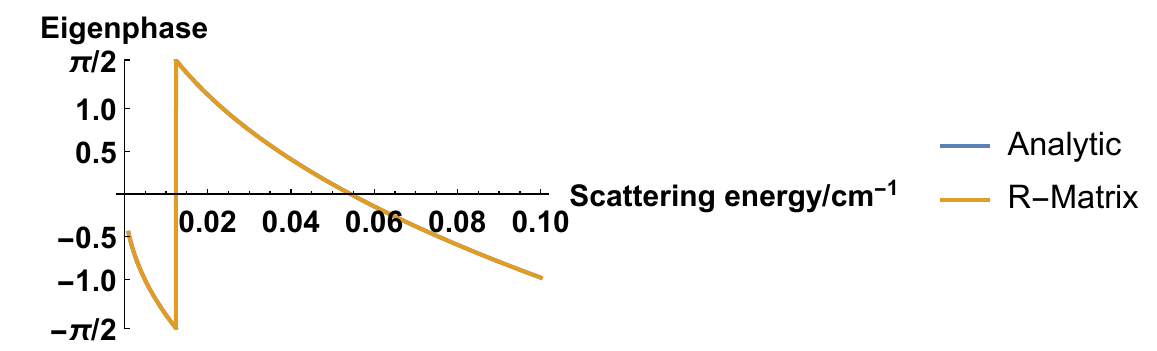}
\includegraphics[scale=0.8]{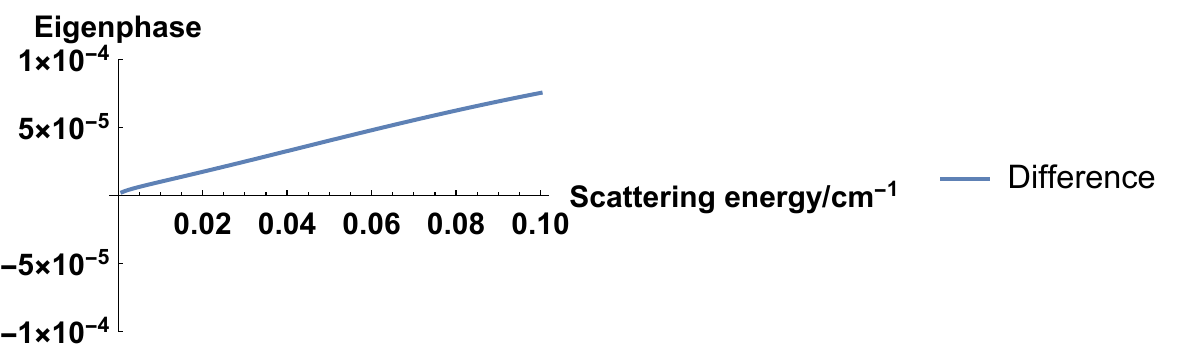}
\caption{Upper plot: eigenphase (in radians) for a scattering event for the
Morse potential of Fig.~\ref{fig:morsepotential2} calculated both analytically
and using R-matrix methodology. The two lines overlap.  Lower plot: difference (analytic $-$ R-matrix)
in eigenphase (in radians) between the two methods.}
\label{fig:EigenphasePlot1}
\end{figure}

Figure~\ref{fig:EigenphasePlot1} compares the RmatReact numerical
eigenphase  to the analytic solution for the
eigenphase given by Eq.~(\ref{eq:AnalyticMorse}) over the scattering energy range of $0.001$ to $0.1$~\cm\ ($0.00144$ to $0.144$ K). 
The root mean square difference between the analytic and numeric
results is approximately $4.6\times 10^{-5}$ radians, which is small.

Analytic and numerical results for scattering length and effective range are presented in Table \ref{tab:ScatLengthEffRange}. Again the results given by the two methods are very similar.

\begin{table}
\renewcommand{\arraystretch}{1.15}
\centering
\caption{Table of comparisons for the analytic and numeric scattering length and effective range. The relative error refers to the difference between each quantity's numeric and analytic values divided by the analytic value (analytic minus numeric, divided by analytic).}
\label{tab:ScatLengthEffRange}
\begin{tabular}{llll} 
\hline\hline
                  & Analytic/\AA & R-matrix/\AA & Relative error  \\ 
\hline
Scattering Length & $10.166078$        & $10.166133$        & $-5.34 \times 10^{-6}$               \\ 
Effective Range   & $1.6537298$        & $1.6667562$        & $-0.00788$                \\
\hline\hline
\end{tabular}
\end{table}

\subsection{Numerical Parameters}
\label{NumericalParameters}

To investigate the accuracy of the R-matrix method in comparison to
the analytic results, the numerical parameters used in the algorithm
were varied and the resultant error was plotted. The seven numerical
parameters which the method relies on are summarised in Table
\ref{tab:NumericalParameters}.

To encapsulate all of the information in the lower plot of
Fig.~(\ref{fig:EigenphasePlot1}) in one number, the error metric used
was the root mean square deviation (RMSD) between the eigenphase,
$\delta(E)$ calculated using the R-matrix method
($\delta_{\rm{num}}(E) $) and the analytic eigenphase
($\delta_{\rm{ana}}(E)$). The eigenphase was calculated for $100$
equally spaced scattering energy values between $0.001$ and $0.1$ \cm
. The error characteristic, the RMSD, was then calculated using:

\begin{equation}
\label{eq:EPRMSD}
\delta_{\rm{RMSD}} =\sum_{i=1}^{100} \sqrt{\frac{(\delta_{\rm{ana}}(E_i) - \delta_{\rm{num}}(E_i))^2}{100}}.
\end{equation}

\begin{figure}[H]
\includegraphics[scale=0.5]{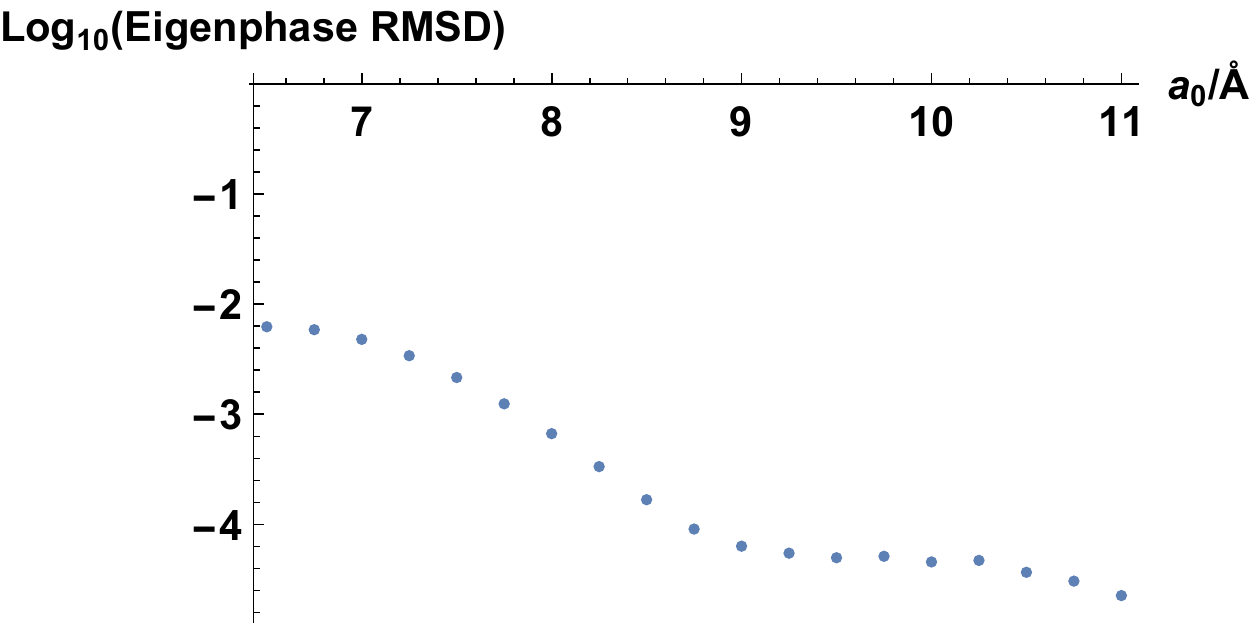}\hspace{1cm}
\includegraphics[scale=0.5]{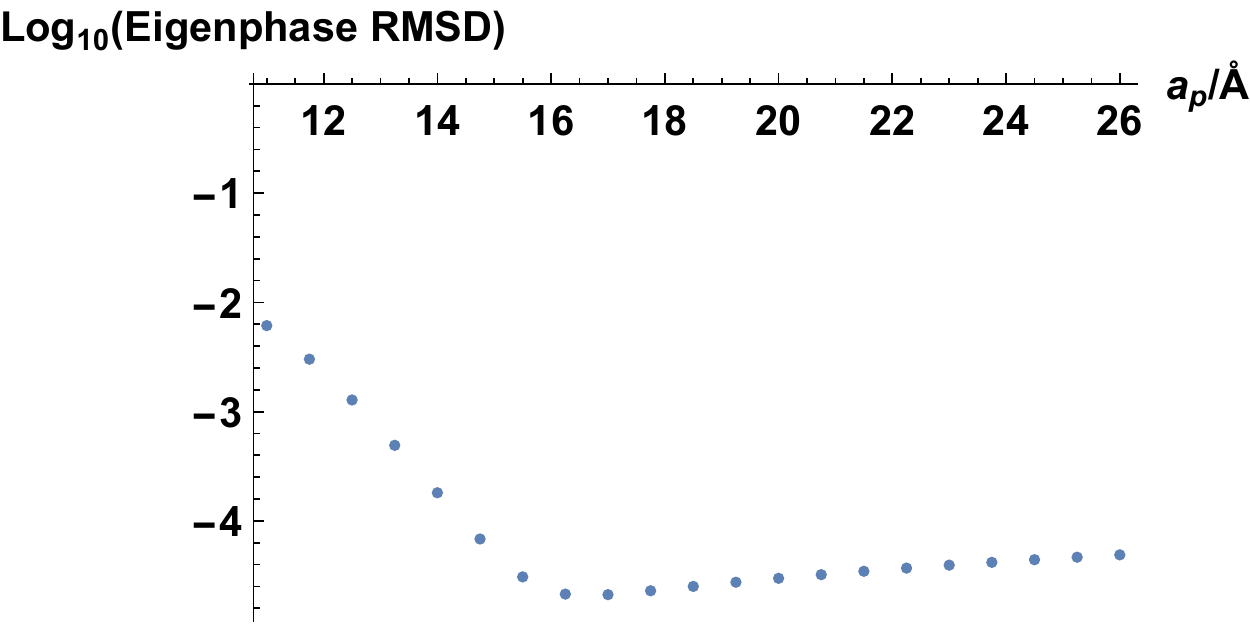}\vspace{0.5cm}
\includegraphics[scale=0.5]{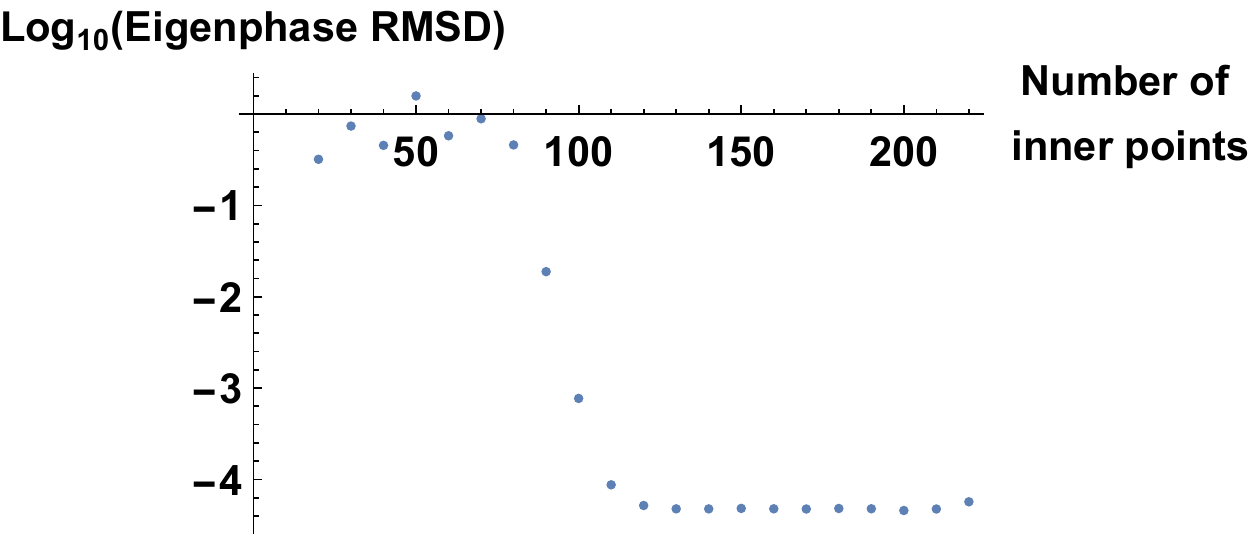}\hspace{1cm}
\includegraphics[scale=0.5]{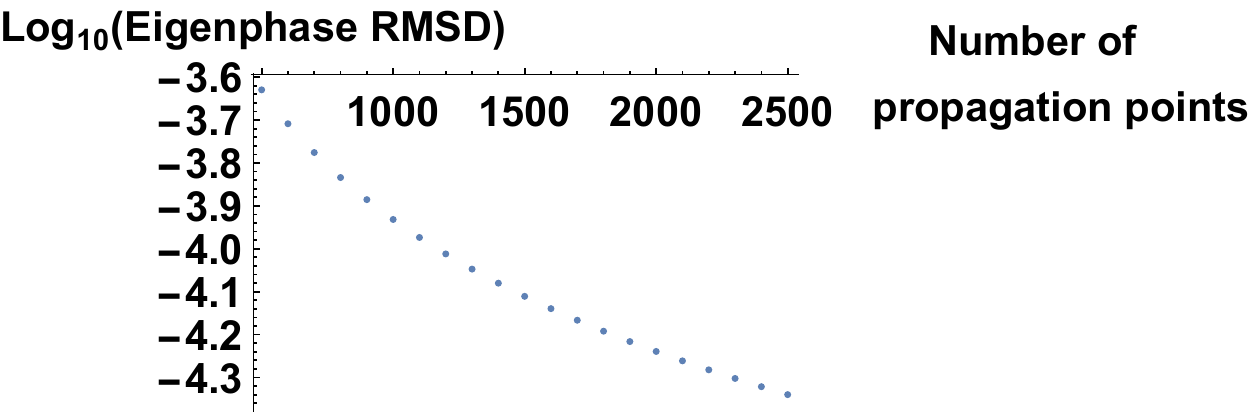}\vspace{0.5cm}
\includegraphics[scale=0.5]{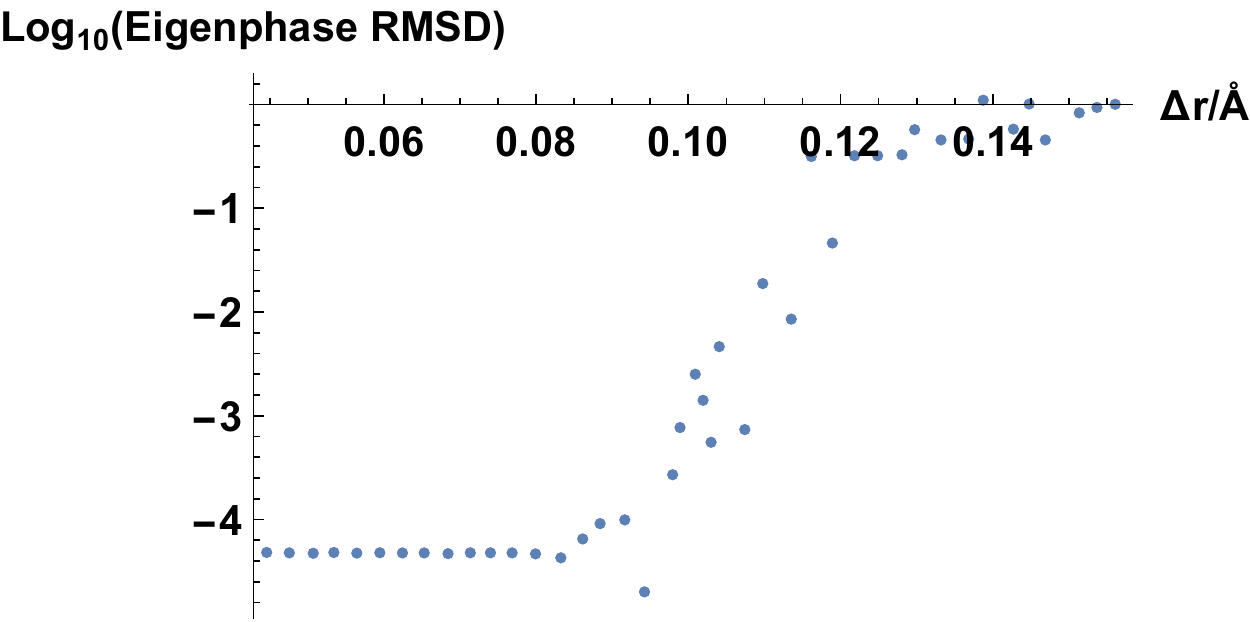}\hspace{1cm}
\includegraphics[scale=0.5]{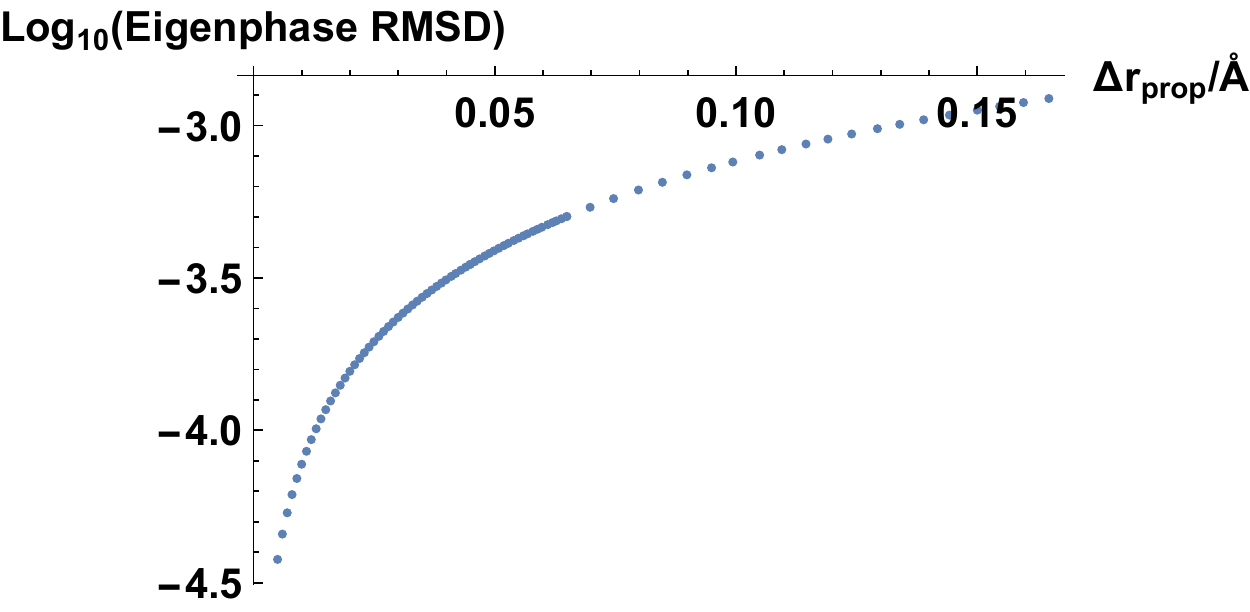}
\caption{Top left: The log of the RMSD of the eigenphase plotted against $a_0$ between $6.5$\AA\ and $11$\AA. The other parameters were held constant at $N=200$, $N_{\rm{prop}}=2500$, $r_{\rm{prop}}=25$\AA.\newline
Top right: The log of the RMSD of the eigenphase plotted against $a_p$ between $11$\AA\ and $26$\AA. The other parameters were held constant at $N=200$, $N_{\rm{prop}}=2500$, $a_0=10$\AA.\newline
Middle left: The log of the RMSD of the eigenphase plotted against $N$ between $20$ and $220$. The other parameters were held constant at $N_{\rm{prop}}=2500$, $a_0=10$\AA, $r_{\rm{prop}}=25$\AA.\newline
Middle right: The log of the RMSD of the eigenphase plotted against $N_{\rm{prop}}$ between $500$ and $2500$. The other parameters were held constant at $N=200$, $a_0=10$\AA, $r_{\rm{prop}}=25$\AA.\newline
Bottom left: The log of the RMSD of the eigenphase plotted against $\Delta r$ between $0.0445982$\AA\ and $0.156094$\AA. $N$ was allowed to vary between $223$ and $63$ to vary $\Delta r$. The other parameters were held constant at $a_0=10$\AA, $N_{\rm{prop}}=2500$, $r_{\rm{prop}}=25$. \newline
Bottom right: The log of the RMSD of the eigenphase plotted against $\Delta r_{\rm{prop}}$ between $0.005$\AA\ and $0.164835$\AA. $N_{\rm{prop}}$ was allowed to vary between $3000$ and $90$ to vary $\Delta r_{\rm{prop}}$. The other parameters were held constant at $N=200$, $a_0=10$\AA, $r_{\rm{prop}}=25$.}
\label{fig:NumericalPlots}
\end{figure}

A version of this error metric which involved (numerically)
integrating the squared difference over the energy range was tested,
and found to give the same results as merely sampling over $100$
equally spaced points in the energy range. Plotting
$\delta_{\rm{RMSD}}$ as a function of different error parameters
facilitated the assessment of the numerical stability of the method.
These plots can be found in Fig.~(\ref{fig:NumericalPlots}). For all
of the plots in Fig.~(\ref{fig:NumericalPlots}), $r_{\rm{min}}$ was
kept constant at $0.01$\AA.

\begin{table}[]
\renewcommand{\arraystretch}{1.2}
\setlength{\tabcolsep}{7pt}
\centering
\caption{Table of numerical parameters}
\label{tab:NumericalParameters}
\begin{tabular}{lll}
\hline\hline
Symbol & Definition                                                             & Units \\ \cline{1-3}
\hline
$N$      & Number of inner region states and grid points                          & Unitless     \\
$N_{\rm{prop}}$ & Number of propagation points                                           & Unitless     \\
$r_{\rm{min}}$  & Start of inner region & $\rm{\AA}$   \\
$a_0$     & End of inner region and start of propagation                                                    & $\rm{\AA}$   \\
$a_p$ & End of propagation                                                     & $\rm{\AA}$   \\
$\Delta r$ & $\frac{a_0 - r_{\rm{min}}}{N-1}$ Average inner region grid spacing & $\rm{\AA}$   \\
$\Delta r_{\rm{prop}}$ & $\frac{a_p -a_0}{N-1}$ Average propagator grid spacing & $\rm{\AA}$   \\
\hline\hline
\end{tabular}
\end{table}

When varying $a_0$, any $a_0$ value above approximately $9$~\AA\
appears to produce converged results where the error changes very
little. This is likely because a value of $a_0$ which is too small
cannot accurately `capture' all of the bound states of the potential
well. Since the final bound state is of the order $10^{-2}$~\cm\ in
depth, $V(a_0)$ must be approximately of that order for the state to
be found by the method.

When varying $a_p$, any value above $16$~\AA\ appears to produce
converged results; however, the error increases slightly as $a_p$
is extended beyond $16$~\AA. This is likely due to $\Delta
r_{\rm{prop}}$ increasing as $N_{\rm{prop}}$ is held constant, which
decreases the accuracy of the approximations made in the propagator
method.

When varying $N$ and $\Delta r$ (where $\Delta r$ is increased by decreasing $N$
and vice-versa), there is a clear point where increasing $N$ further
has no effect, but where decreasing $N$ even slightly significantly
increases the error. This suggests that the method is converging on a
solution once the grid spacing is sufficiently small, as is common in
numerical integration techniques. This further suggests that this solution's RMSD from
the analytic solution is approximately $10^{-4}$.

Finally, when varying $N_{\rm{prop}}$ and $\Delta r_{\rm{prop}}$, the
method appears to produce results with very low error for all values
of $N_{\rm{prop}}$ and $\Delta r_{\rm{prop}}$ tested, with only slight
variation in the error recorded. This suggests that it is possible to
propagate the R-matrix using very few, very wide steps and still
produce accurate results. However, this may be a consequence of using as the test potential the
Morse oscillator potential, since it decreases exponentially with distance and
thus varies very little in the outer region. More relatistic potentials are longer-range and multipolar in nature at large $r$, so narrower steps may be needed in the propagation.

\section{Conclusions and outlook}
\label{Conclusions}

We clearly demonstrate that we can obtain excellent results using our R-matrix implementation for low-energy
scattering within a Morse oscillator potential. Asymptotically this potential decays exponentially, which makes it unlike physical
potentials, which have a much longer range. Physical potentials decay as $r^{-n}$, where $n$ is a positive integer. 

The next step is to implement DVR shape functions into variational
nuclear motion codes to facilitate the calculation of boundary
amplitudes within these codes. This been done for the general diatomic
code Duo \cite{jt609} and triatomic code DVR3D \cite{jt338}. The
diatomic problems for which tests have been run so far all involve a
single asymptotic channel, which makes R-matrix propagation
straightforward. In general this will not be true and it will be
necessary to consider multichannel problems. To address this issue we
have successfully performed propagations with a general code
originally designed for electron -- atom problems \cite{farm}.  This
code now needs generalising to provide automated resonance fitting
\cite{jt31,jt651} and bound state finding \cite{jt106} features.

We intend to use this new methodology on physical problems, and to create a generalisation of the R-matrix
formalism to allow the explicit treatment of reactive processes. We have
conducted preliminary tests similar to the ones presented here on more accurate Ar -- Ar potentials with multipolar long-range expansions, for which the leading term is $n=6$, and also obtained excellent results. All of these results will be reported elsewhere.

\section*{Acknowledgments}
This project has received funding from the European Union's Horizon
2020 research and innovation programme under the Marie
Sklodowska-Curie grant agreement No 701962 and from the EPSRC.

\bibliographystyle{spphys}

\end{document}